\documentclass[12 pt]{article}

\usepackage{graphicx}

\newcommand{\approxge}{\stackrel{>}{\scriptstyle\sim}}
\newcommand{\approxle}{\stackrel{<}{\scriptstyle\sim}}
\newcommand{\cbar}{\overline{c}}
\newcommand{\qbar}{\overline{q}}
\newcommand{\pbar}{\overline{p}}
\newcommand{\bbar}{\overline{b}}

\begin{document}\setlength{\unitlength}{1mm}

\begin{flushright}
\begin{tabular}{l}
AMES-HET-98-07 \\
MADPH-98-1052 \\
hep-ph/9806486 \\
June 1998
\end{tabular}
\end{flushright}
\begin{center}
\vspace{12pt} 
{\LARGE Single Top Quark Production \\
via FCNC Couplings at Hadron Colliders} \\
\vspace{12pt} T. Han$^a$, M. Hosch$^b$, K. Whisnant$^b$,
Bing-Lin Young$^b$, and X. Zhang$^c$ \\ 
{\it $^a$ Department of Physics, University of Wisconsin, \\
Madison, WI  53706, USA\\
$^b$ Department of Physics and Astronomy, Iowa State University,\\
Ames, Iowa  50011, USA \\
$^c$ CCAST (World Laboratory), P.O. Box 8730, Beijing 100080, and\\
Institute of High Energy Physics, Academia Sinica, Beijing 100039, China\\}
\vspace{120pt}
\end{center}

\begin{abstract}

We calculate single top--quark production at hadron colliders via the
chromo--magnetic flavor--changing neutral current couplings $\bar tcg$
and $\bar tug$. We find that the strength for the anomalous $\bar tcg$
($\bar tug$) coupling may be probed to $\kappa_c / \Lambda =
0.092\rm{~TeV}^{-1}$ ($\kappa_u / \Lambda = 0.026 \rm{~TeV}^{-1}$) at
the Tevatron with $2 \rm{~fb}^{-1}$ of data and $\kappa_c / \Lambda =
0.013\rm{~TeV}^{-1}$ ($\kappa_u / \Lambda = 0.0061 \rm{~TeV}^{-1}$) at
the LHC with $10 \rm{~fb}^{-1}$ of data. The two couplings may be
distinguished by a comparision of the single top signal with the direct
top and top decay signals for these couplings.

\end{abstract}

\pagebreak

\section{Introduction}

Since the discovery of the top quark at the Fermilab
Tevatron~\cite{cdfd0}, there has been considerable interest in exploring
the properties of the top quark. Its unusually large mass close to the
electroweak symmetry--breaking scale makes it a good candidate for
probing for physics beyond the Standard Model (SM)~\cite{heavyt}, and
has also given rise to explorations of anomalous couplings of the top
quark~\cite{exprmnts,Rizzox,Han1x}.

A distinctive set of anomalous interactions is given by the 
flavor--changing chromo--magnetic operators~\cite{Han2x,Hoschx}:
\begin{equation}
\frac{\kappa_f}{\Lambda} g_s \overline{f} \sigma^{\mu \nu}
\frac{\lambda^a}{2} t G^a_{\mu \nu}\ +\ h.c.\ ,
\label{anom}
\end{equation}
where $\Lambda$ is the new physics scale, 
$f=u$ or $c$, the $\kappa_f$ define the strength of the 
$\overline{t}ug$ or $\overline{t}cg$ couplings,
and $G^a_{\mu \nu}$ is the gauge field tensor of the gluon.
Although such interactions can be produced by higher order radiative
corrections in the SM, the effect is too small to be observable
\cite{Eilamx}. Any signal indicating these
types of couplings is therefore evidence of physics beyond the SM, and
will shed more light on flavor physics in the top--quark sector.
It has been suggested that couplings of this type may be large in some
extensions to the standard model, especially in models with multiple
Higgs doublets such as supersymmetry~\cite{Eilamx,Grzadx,yyz}.
Models with new dynamical interactions of the top quark~\cite{Hillx} 
and those in which the top quark is composite~\cite{Georgix} or has a 
soliton~\cite{Zhangx} structure could also give rise to
the new couplings in Eq.~(\ref{anom}).

Currently, there only exist rather loose bounds on these
anomalous couplings. The good agreement between the SM theory and
experiments in top--quark production at the Fermilab
Tevatron~\cite{cdfd0} places a modest limit on the couplings in
Eq.~(\ref{anom}). By looking for signals from $t \to cg$ or $ug$
decays~\cite{Han2x}, the coupling parameters $\kappa_f / \Lambda$ can be
constrained down to $0.43 \rm{~TeV}^{-1}$ with the existing 
Tevatron Run 1 data.

In this paper, we examine the effect of these couplings on single
top--quark production at the Tevatron and the CERN LHC. There are four
different subprocesses which lead to one top quark in
the final state together with one associated jet. The identity of the
associated jet depends upon the initial state of the system: 
\begin{eqnarray}
q \qbar \to t \cbar,\ \ 
g g \to t \cbar,\ \ 
c q(\qbar) \to t q(\qbar),\ \ 
c g \to t g.
\label{process}
\end{eqnarray}
We also consider similar processes which replace the $c$ quark 
with the $u$ quark, as well as similar processes for single anti--top
quark production. The advantage of this class of events 
is the unique production mechanism and the distinctive 
final state kinematics over the SM backgrounds.

The effect of the anomalous $\bar tcg$ coupling on single top quark
production via the $q\bar q$ process in Eq.~(\ref{process}) at the
Tevatron has been studied in Ref.~\cite{Malkawix}. It was found that
$\kappa_c / \Lambda$ can be measured to $0.4\ \rm{TeV}^{-1}$ with the
existing Tevatron data, comparable to the limit obtained by the
anomalous top decay processes \cite{Han2x} mentioned earlier. We find,
however, that the other processes in Eq.~(\ref{process}) are also
important, especially at higher energies. This is particularly true for
the case of $\kappa_u/\Lambda$.

This paper is organized as follows. In Sec.~2, we discuss the
signal and background calculations. We also study a set of acceptance
cuts to optimize the signal observability. In Sec.~3, we discuss
our results and related issues, and make some concluding
remarks.

\section{Signal and Background Calculations}

We calculate the tree level cross sections for single top 
(or anti--top) quark production of Eq.~(\ref{process}) 
at hadron colliders using the flavor--changing chromo--magnetic 
couplings in Eq.~(\ref{anom}). For completeness, we study the
following collider parameters for the center--of--mass energy and
integrated luminosity:
\begin{table}[h]
\begin{center}
\begin{tabular}{lccc}
         &       &  $E_{cm}$ (TeV)  & ${\cal L}$ (fb$^{-1}$)\\
Tevatron & Run 1 &  1.8             & 0.1 \\ 
         & Run 2 &  2               & 2   \\ 
         & Run 3 &  2               & 30  \\ 
LHC      &       & 14               & 10  \\ 
\end{tabular}
\end{center}
\end{table}

\begin{figure}[tbh]
\begin{center}
\includegraphics{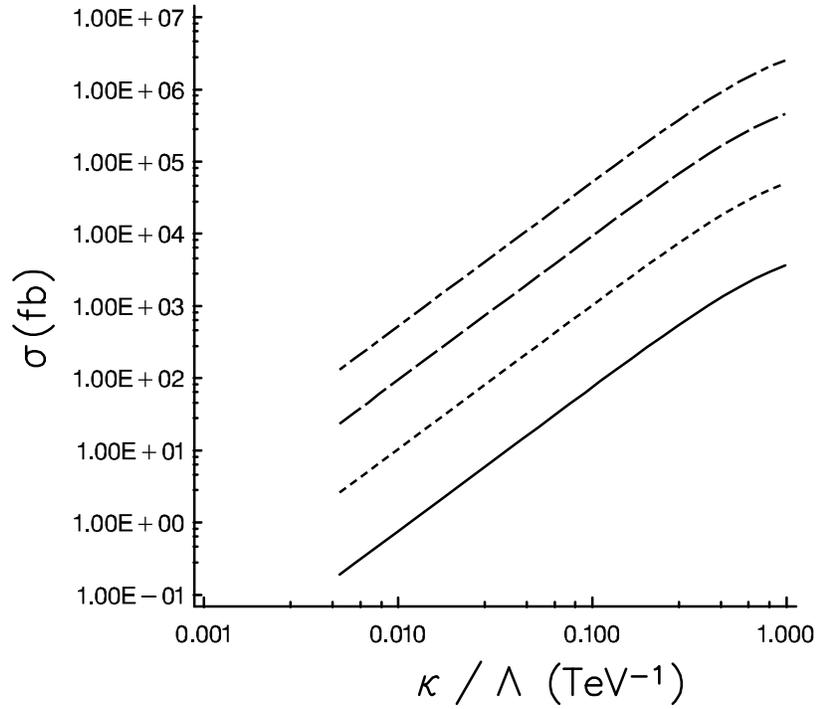}
\end{center}
\caption{\label{scalingx}
Cross sections for single top--quark production $pp (\bar p) \to tj$
versus $\kappa_f / \Lambda$ at the Tevatron and LHC. The solid and short
dashed lines are for the $\bar tcg$ and $\bar tug$ coupling at the
Tevatron, respectively.  The long dashed and dash--dotted lines are at
the LHC.}
\end{figure}

As for the final state signature, while the $t \to cg$ or $ug$ decay
will occur due to the anomalous couplings in Eq.~(\ref{anom}), the
branching ratio becomes negligible when $\kappa_f/\Lambda$ is less than
about $0.2 \rm{~TeV}^{-1}$. For this reason, and since the top--quark
decay $t \to b W \to b \ell \nu_\ell$ is easier to identify than the
pure hadronic mode, we will choose as the search mode for the signal
\begin{equation}
p \pbar \to t + j \to b \ell \nu_\ell + j \,,
\label{signal}
\end{equation}
where $j$ is a light parton jet and $\ell = e,\mu$.
We assume that the top quark is on mass shell when we 
calculate the decay process, but the spin correlation
effects are fully incorporated. 
In determining the branching ratio of 
$ t \to Wb$, we have properly included the contribution 
from $t \to cg$ or $ug$.  This term is proportional to 
$|\kappa_f/\Lambda|^2$ and it only appreciably affects the
branching ratio if $\kappa_f/\Lambda \approxge 0.2 \rm{~TeV}^{-1}$.
The calculated cross sections are shown in Fig.~\ref{scalingx}
as a function of $\kappa_f / \Lambda$,
for the Fermilab Tevatron ($p \pbar$ at the
center of mass energy $\sqrt{s}=1.8$ TeV) 
and for the CERN LHC ($p p$ at $\sqrt{s}=14$ TeV),
using the MRSA structure functions \cite{MRSx}.
Results for the Tevatron at 2 TeV are nearly 
indistinguishable from the cross sections shown here
for 1.8 TeV.

In order to simulate the detector effects in identifying the signal,
we made a series of standard cuts on the transverse momentum $p_T^{}$,
the missing transverse momentum resulting from the neutrino $p\llap/_T$,
pseudorapidity $\eta$, and the jet (lepton) separation $\Delta R$. We
call these the ``basic cuts'':
\begin{eqnarray}
p_{Tb},p_{Tj},p_{Tl},p\llap/_T & \ge & 15 \rm{~GeV}\ , 
\label{basic1} \\
\eta_b,\eta_j,\eta_l & \le & 2.5\ , 
\label{basic2} \\
\Delta R_{jj},\Delta R_{jl} & \ge & 0.4\ .
\label{basic3}
\end{eqnarray}
To be more realistic, we also assume a Gaussian
smearing of the energy deposited in calorimeters, given by:
\begin{eqnarray}
\Delta E / E & = & 30 \% / \sqrt{E} \oplus 1 \% \rm{,~for~leptons~,} \\
             & = & 80 \% / \sqrt{E} \oplus 5 \% \rm{,~for~hadrons~,} 
\end{eqnarray}
where $\oplus$ indicates that the two terms are added in quadrature.

The major source of background to signal production in 
Eq.~(\ref{signal}) is 
\begin{equation}
p \pbar \to W + j j \,,  
\end{equation}
where the jets are light quarks or gluons.  An effective way to reduce
the background is to employ $b$--tagging. We take a $b$--tagging
efficiency of $36\%$ at Run 1 of the Tevatron, and $60\%$ at Runs 2 and
3 and at the LHC. We also assume that $1\%$ of non--$b$ quarks would be
misidentified as $b$ quarks in all of the experiments, which gives a
suppression of the $W + jj$ background by a factor of $\approx
50$.  Other significant backgrounds are standard model single top quark
production where one light quark accompanies the top quark in the final
state \cite{willen}, and standard model $Wb\bar b$ production. 
Backgrounds with two
$b$ quarks in the final state will mimic our signal if one of
the $b$ quarks is missed by the $b$ tag.

We used the VECBOS Monte Carlo~\cite{VECBOSx} to calculate the cross
sections for the $W jj$ and $W b \bbar$ backgrounds. We made two
modifications to the VECBOS code. First, we added a routine to calculate
some of the kinematic distributions, as described below, that we found
to be very helpful for background suppression. Second, since our signal
process involves one $b$ quark in the final state, $b$--tagging was
crucial in eliminating the $Wjj$ background; we modified the code to
randomly choose one of the jets to be mis--identified as a $b$--jet
(with probability of 1\%).  For the calculation of the standard model
single top processes $bq \rightarrow tq$ and $q\bar q \rightarrow t \bar
b$ we used our own monte carlo routine~\cite{thesis}.

To further enhance the signal relative to the $Wjj$ and $Wb\bar b$
backgrounds, we imposed a constraint on $M_{bW}$, the
invariant mass of the $W$ and $b$ quark,
which should be peaked at $m_t$ for the signal.  To
experimentally determine $M_{bW}$, one must reconstruct $p_t = p_b + p_l
+ p_{\nu_l}$.  The neutrino is not observed, but its transverse momentum
can be deduced from the missing transverse momentum.  The longitudinal
component of the neutrino momentum is determined by setting $M_{l\nu_l}
= M_W$, and is given by:
\begin{equation}
p^{\nu_l}_L = \frac{\chi p^l_L \pm \sqrt{ \vec{p}_l^{\, 2} (\chi^2 -
p_{Tl}^2 p_{T\nu_l}^2) }}{p_{Tl}^2}~,
\end{equation}
where
\begin{equation}
\chi = \frac{M_W^2}{2} + \vec{p}^{\, l}_T \cdot \vec{p}^{\nu_l}_T ~,
\end{equation}
and $p_L$ refers to the longitudinal momentum. Note that there is a two
fold ambiguity in this determination.  We chose the solution in which
$M_{bW}$ is closer to the mass of the top quark.
This process artificially inserts a broad peak in the background at
$M_{bW} = m_t$, but since the signal peak is much sharper, the $M_{bW}$
distribution is still an effective variable to use to increase the
signal--to--background ratio.

\begin{figure}[p]
\begin{center}
\includegraphics{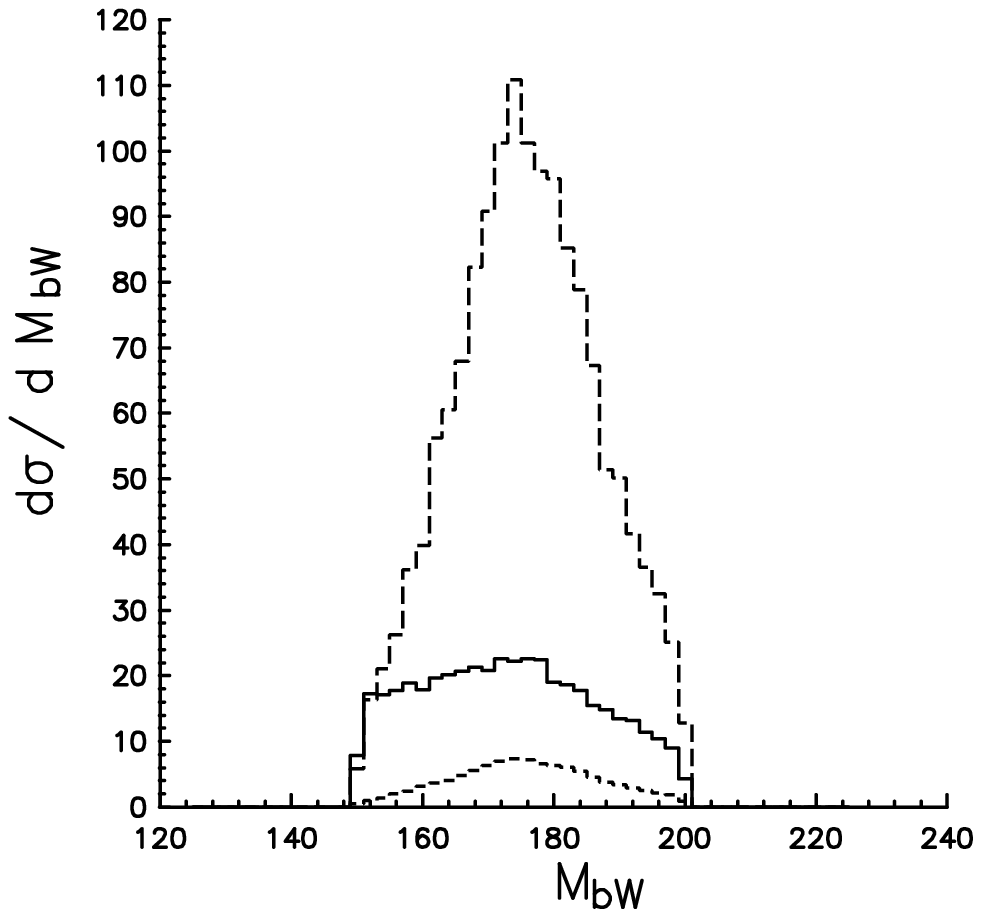}
\end{center}
\caption{\label{dist1x} Distributions for $M_{bW}$, in fb/GeV, after
the cuts in Eqs.~(\ref{basic1})--(\ref{basic3}) and
(\ref{mbwcut})--(\ref{delrjjcut}) and $b$--tagging at the Tevatron
Run~1. The solid line represents the sum of all of the background
processes, the long dashed line is the sum of the up--quark signal
processes, and the short dashed line is the sum of the charm--quark
signal processes, with $\kappa_f/\Lambda = 0.2$~TeV$^{-1}$.}
\end{figure}

\begin{figure}[p]
\begin{center}
\includegraphics{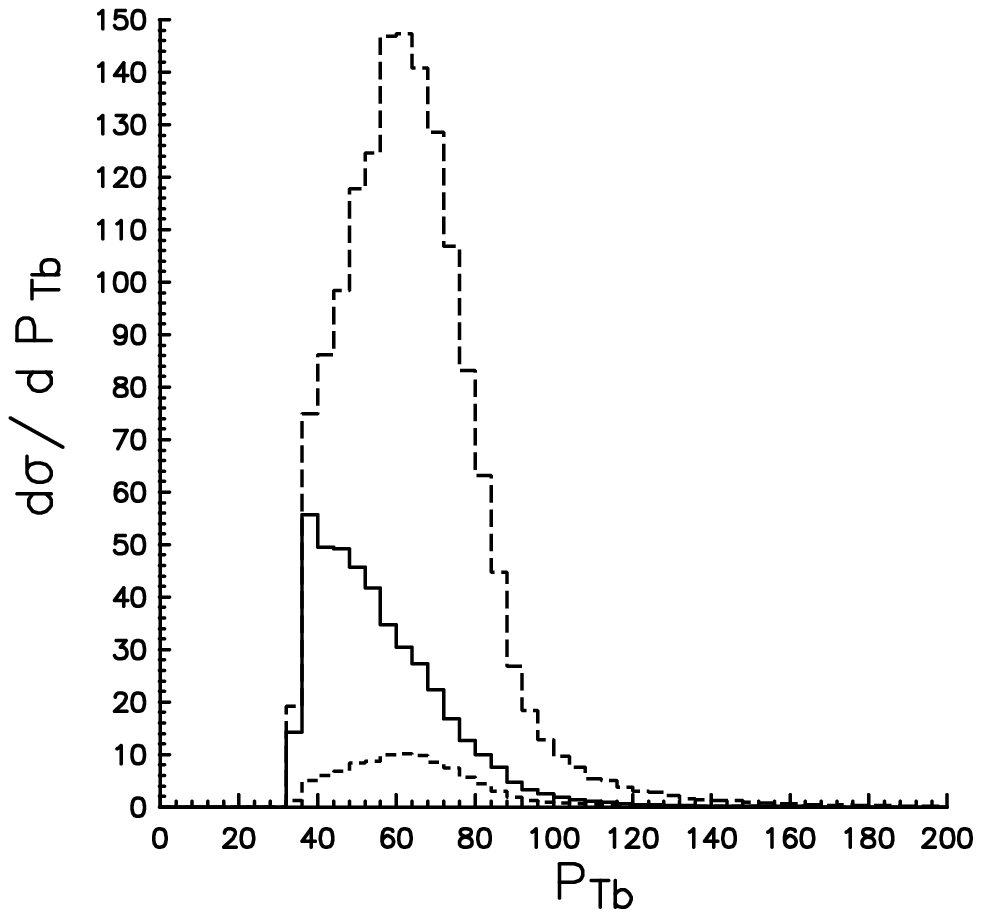}
\end{center}
\caption{\label{dist2x} Distributions for $p_{Tb}$, in fb/GeV, after
the cuts in Eqs.~(\ref{basic1})--(\ref{basic3}) and
(\ref{mbwcut})--(\ref{delrjjcut}) and $b$--tagging at the Tevatron
Run~1. The solid line represents the sum of all of the background
processes, the long dashed line is the sum of the up--quark signal
processes, and the short dashed line is the sum of the charm--quark
signal processes, with $\kappa_f/\Lambda = 0.2$~TeV$^{-1}$.}
\end{figure}

\begin{figure}[p]
\begin{center}
\includegraphics{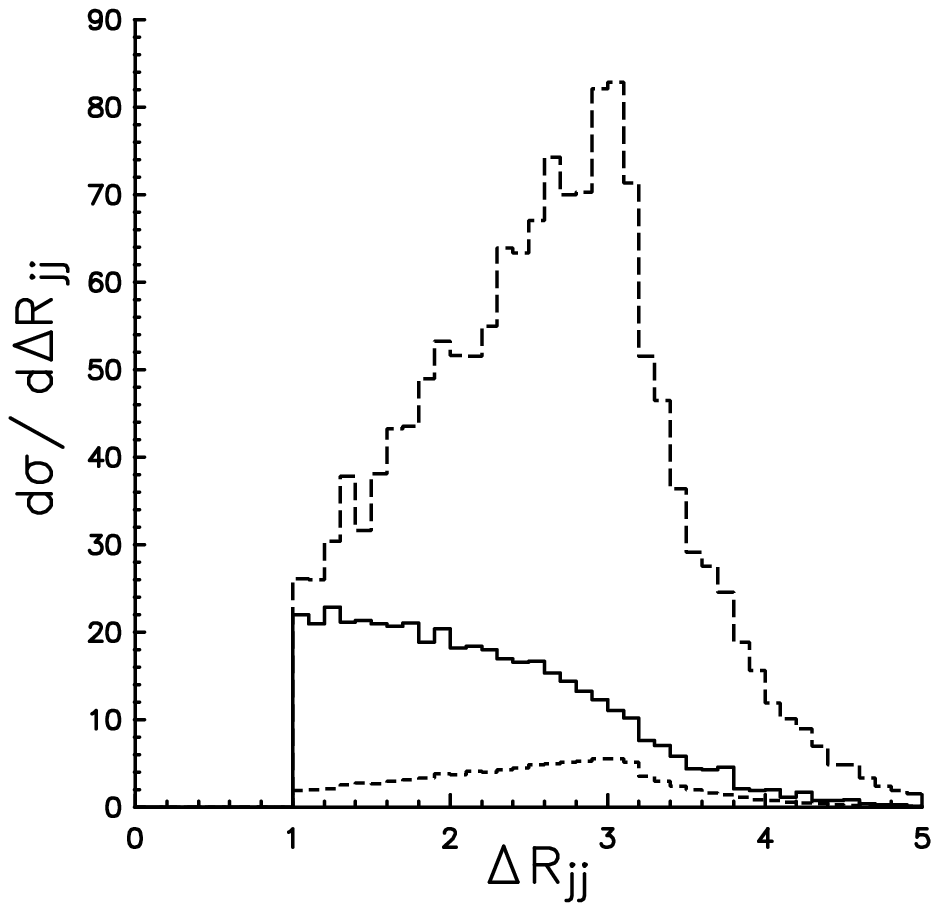}
\end{center}
\caption{\label{dist3x} Distributions for $\Delta R_{jj}$, in fb
per unit $\Delta R_{jj}$, after
the cuts in Eqs.~(\ref{basic1})--(\ref{basic3}) and
(\ref{mbwcut})--(\ref{delrjjcut}) and $b$--tagging at the Tevatron
Run~1. The solid line represents the sum of all of the background
processes, the long dashed line is the sum of the up--quark signal
processes, and the short dashed line is the sum of the charm--quark
signal processes, with $\kappa_f/\Lambda = 0.2$~TeV$^{-1}$.}
\end{figure}

\begin{figure}[p]
\begin{center}
\includegraphics{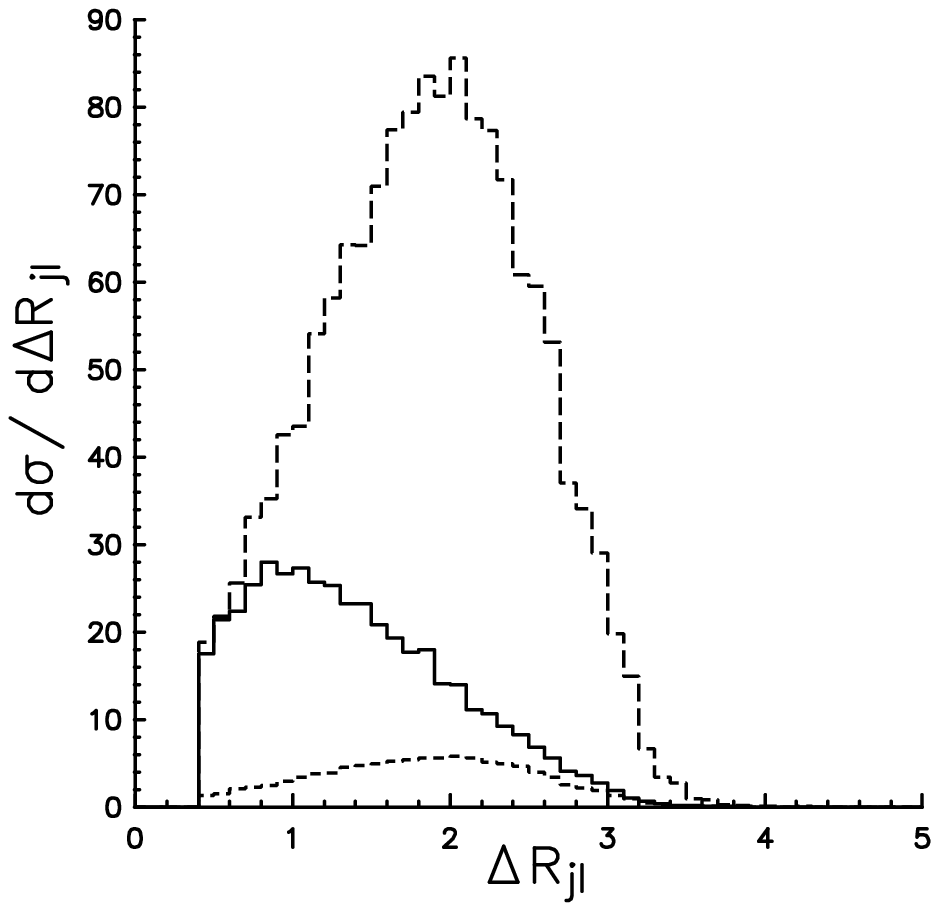}
\end{center}
\caption{\label{dist4x} Distributions for $\Delta R_{jl}$, in fb
per unit $\Delta R_{jl}$, after
the cuts in Eqs.~(\ref{basic1})--(\ref{basic3}) and
(\ref{mbwcut})--(\ref{delrjjcut}) and $b$--tagging at the Tevatron
Run~1. The solid line represents the sum of all of the background
processes, the long dashed line is the sum of the up--quark signal
processes, and the short dashed line is the sum of the charm--quark
signal processes, with $\kappa_f/\Lambda = 0.2$~TeV$^{-1}$.}
\end{figure}

To find a way to further isolate the signal from the background, we
examined the kinematic distributions in $\sqrt{\hat{s}}$, $M_{bW}$,
$p_T$, $\eta$, and $\Delta R$. Three of the variables, $M_{bW}$,
$p_{Tb}$, and $\Delta R_{jj}$, are especially useful for significantly
suppressing the background and therefore improving the discovery limit
for the $\kappa_f/\Lambda$. Just as $M_{bW}$ has a peak near $m_t$ for
the signal, $p_{Tb}$ develops its Jacobian peak near
$\frac{1}{2}m_t\sqrt{1-M_W^2/m_t^2}$.  Furthermore, $\Delta R_{jj}$
reaches a peak near $\pi$ for the signal since the two jets are largely
back-to-back.  To effectively reduce the background versus the signal we
applied the additional cuts for the Tevatron Run~1
\begin{eqnarray}
150 {\rm~GeV} &\leq& M_{bW} \leq 200 {\rm~GeV} \,,
\label{mbwcut}\\
p_{Tb} &\geq& 35 {\rm~GeV} \,,
\label{ptbcut}\\
\Delta R_{jj} &\geq& 1.0 \,, \ 
\Delta R_{lj} \geq 0.4 \,.
\label{delrjjcut}
\end{eqnarray}
The distributions for these variables are shown for Run~1 in
Figs.~\ref{dist1x}--\ref{dist4x} after the cuts in
Eqs.~(\ref{basic1})--(\ref{basic3}) and
(\ref{mbwcut})--(\ref{delrjjcut}) are applied. For Run~2 and Run~3 and
for the LHC, the signal--to--background ratio can be improved by
changing the cuts in Eqs.~(\ref{mbwcut})--(\ref{delrjjcut}). The
optimized cuts for each collider option are shown in Table~\ref{cutsx}.
The cross sections for the signal and background channels after all cuts
are shown in Table~\ref{ratesx}.

In Ref.~\cite{Malkawix} anomalous single top quark production was
studied through only one channel, $q \qbar \to t \cbar$.  As
Table~\ref{ratesx} shows, this is the least important of all of the
channels in Eq.~(\ref{process}). While it would seem that the presence
of initial state valence quarks ought to make this the dominant process,
the massless $t$--channel exchange of a gluon in the $c g \to t g$
process more than makes up for the lack of initial state valence quarks,
and it becomes the most important process.  Each of the other processes,
$c q \to t q$ and $g g \to t \cbar$, also have massless, or nearly
massless, $t$--channel exchanges which increase their parton cross
sections.  We also note that in Ref.~\cite{Malkawix} a cut on the center
of mass energy, $\sqrt{\hat{s}} > 300 \rm{~GeV}$, was imposed for the
purpose of reducing the background relative to the signal.  Because of
the dominance of the massless $t$--channel exchanges, the parton cross
section is peaked at lower values of $\sqrt{\hat{s}}$ when all of the
processes in Eq.~(\ref{process}) are included. Hence, the
$\sqrt{\hat{s}}$ cut is not useful in our calculation as it will reduce
the signal drastically.

\section{Discussion and Summary}

We may use the results of the signal and background calculation to
determine the minimum values of $\kappa_c/\Lambda$ and
$\kappa_u/\Lambda$ that can be observed at hadron
colliders. We use the criterion that $N_S \geq 3
\sqrt{N_S+N_B}$ approximately corresponding to a $95 \%$ confidence 
level, where $N_S$ and $N_B$ are the number of signal and
background events, respectively. Since the signal is quadratic in
$\kappa_f/\Lambda$ and we have calculated the signal for $\kappa_f/
\Lambda = 0.2\rm{~TeV}^{-1}$, the minimum value of $\kappa_f/ \Lambda$
is then given by
\begin{equation}
{\kappa_f \over \Lambda} = 0.2 \rm{~TeV}^{-1} \sqrt{{9 \over 2} (1 +
\sqrt{1+{4\over9} \mathcal{L} \sigma_B}) \over \mathcal{L} \sigma_0} \,,
\end{equation}
where $\mathcal{L}$ is the integrated luminosity, $\sigma_B$ is the
total cross section for all of the background processes, and $\sigma_0$
is the cross section for the signal processes evaluated at $\kappa_f /
\Lambda = 0.2\ \rm{TeV}^{-1}$. The discovery limits, calculated using
the basic cuts, the optimized cuts in Table~\ref{cutsx}, and
$b$--tagging, are shown in Table~\ref{limitsx}.

A search for direct top quark production (where the top quark is the
only particle in the final state of the parton process) will also place
a limit on the size of $\kappa_f/\Lambda$~\cite{Hoschx}. This process
relies on its rare signature and on the large fraction of gluons in the
initial state to boost its signal relative to the background. The up
quark operator in Eq.~(\ref{anom}) has the additional bonus that the
initial state quark is a valence quark. With $b$--tagging (of the top
quark decay products), direct top quark production provides nearly ideal
conditions for measuring the anomalous coupling parameters.  Using this
process, $\kappa_c / \Lambda$ ($\kappa_u / \Lambda$) can be measured
down to $.062 \rm{~TeV}^{-1}$ ($.019 \rm{~TeV}^{-1}$) at Run 2 of the
Tevatron with $2 \rm{~fb}^{-1}$ integrated luminosity, and to $.0084
\rm{~TeV}^{-1}$ ($.0033 \rm{~TeV}^{-1}$) at the LHC with $10
\rm{~fb}^{-1}$ integrated luminosity. Thus direct top production
provides a second, independent measurement of these anomalous couplings.

\begin{figure}[p]
\begin{center}
\includegraphics{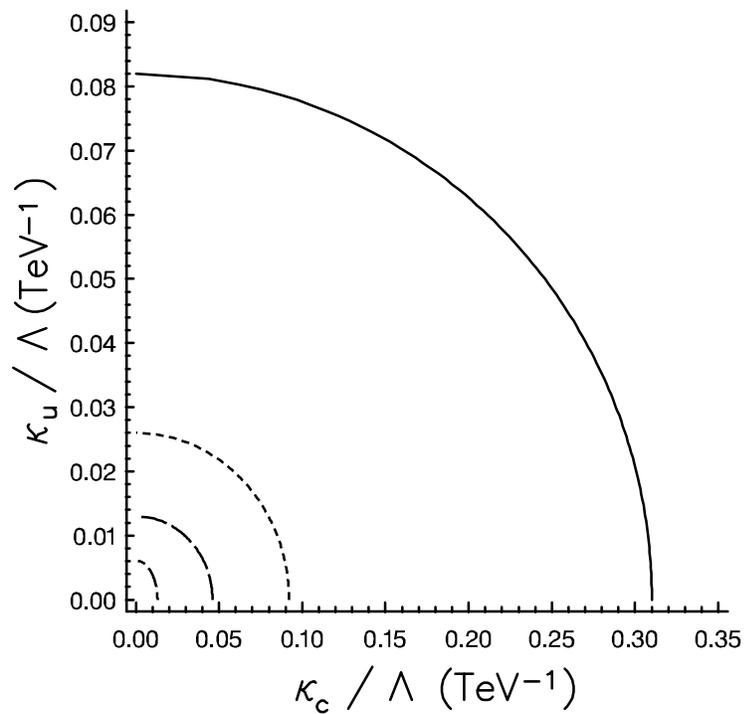}
\end{center}
\caption{\label{2dx} Discovery limits for $\kappa_c / \Lambda$
versus $\kappa_u / \Lambda$ for each of the collider options considered.
The solid, short dashed, and long dashed lines are at Runs 1, 2, and 3 at
the Tevatron respectively.  The dash--dotted line is at the LHC.}
\end{figure}

The $c$ and $u$ quarks have, so far, been treated as if only one of the
couplings exists at a time. If the two anomalous couplings co--exist, we
may simply add the cross sections of the two different couplings
together, since we have treated them in exactly the same manner.  A plot
of their discovery limits, when considered together, is shown in
Fig.~\ref{2dx}.

If a signal is seen, how can we determine if it is due to $\bar tcg$,
$\bar tug$, or perhaps a mixture of the two? If the $c$ or $u$ quark is
in the final state, then charm tagging could in principle distinguish
between the two couplings. However, the signal processes in which this
occurs (i.e., the single top processes $q\bar q \rightarrow t\bar c,
t\bar u$ and $gg \rightarrow t\bar c, t\bar u$, and the anomalous top
decays $t \rightarrow cg, ug$) have smaller cross sections than the
signal processes where the $c$ or $u$ quark is in the initial state (the
single top processes $cq, uq \rightarrow tq$ and $gc, gu \rightarrow
gt$, and the direct top processes $gc, gu \rightarrow t$).  Furthermore,
the efficiency for charm-quark tagging is expected to be low compared to
$b$-quark tagging. Therefore it will be difficult to distinguish the
$\bar tcg$ and $\bar tug$ couplings with charm tagging.

Another possibility is to compare the relative size (after all the cuts
and $b$-tagging) of the anomalous single top cross section
$\sigma_{st}$, the direct top cross section $\sigma_{dt}$~\cite{Hoschx},
and the $t\bar t$ production cross section when the $t$ ($\bar t$)
undergoes an anomalous decay into $cg$ ($\bar cg$) or $ug$ ($\bar ug$),
$\sigma_{at\bar t}$~\cite{Han2x}. For example, at the Tevatron Run~2 for
$\kappa_c/\Lambda = 0.2$~TeV$^{-1}$ these cross sections are
$\sigma_{st}=154$~fb, $\sigma_{dt}=281$~fb, and $\sigma_{at\bar t}=6$~fb,
while for $\kappa_u/\Lambda = 0.2$~TeV$^{-1}$ they are
$\sigma_{st}=1976$~fb, $\sigma_{dt}=2990$~fb, and $\sigma_{at\bar
t}=6$~fb. The ratios $\sigma_{st}/\sigma_{at\bar t}$ and
$\sigma_{dt}/\sigma_{at\bar t}$ are much larger for the $\bar tug$
coupling than for $\bar tcg$. Also, $\sigma_{st}/\sigma_{dt}$ is
somewhat larger for the $\bar tug$ coupling than for $\bar tcg$. Hence,
a comparison of two or more of these signals may help determine whether
the anomalous coupling is $\bar tcg$, $\bar tug$, or a mixture of the
two.

In summary, we calculated the single top--quark production at hadron
colliders via the chromo--magnetic flavor--changing neutral current
couplings $\bar tcg$ and $\bar tug$. We find that the strength for the
anomalous coupling $\bar tcg$ may be probed to $\kappa_c / \Lambda =
0.092 \rm{~TeV}^{-1}$ at the Tevatron with $2 \rm{~fb}^{-1}$ of data at
2~TeV and $\kappa_c / \Lambda = 0.013\rm{~TeV}^{-1}$ at the LHC with
$10 \rm{~fb}^{-1}$ of data at 14~TeV. Similarly, the anomalous coupling
$\bar tug$ may be probed to $\kappa_u / \Lambda = 0.026 \rm{~TeV}^{-1}$
at the Tevatron and $\kappa_u / \Lambda = 0.0061 \rm{~TeV}^{-1}$ at the
LHC.  Assuming $\kappa_c \equiv 1$ ($\kappa_u \equiv 1$), the scale of
new physics $\Lambda$ can be probed to 11~TeV (38~TeV) at the Tevatron
and to 77~TeV (164~TeV) at the LHC.

\section{Acknowledgments}

This work was supported in part by the U.S.~Department of Energy under
Contracts DE-FG02-94ER40817 and DE-FG02-95ER40896. Further support
for T.H. was provided by the University of Wisconsin Research Committee, 
with funds granted by the Wisconsin Alumni Research Foundation.
M. Hosch was also partially supported by GAANN.

\newpage

\begin{table}
\caption{\label{cutsx} Optimized cuts for the discovery of $\kappa_f \over
\Lambda$ with single top quark production for several collider options.}
\vspace{0.2cm}
\begin{center}
\begin{tabular*}{\textwidth}{l@{\extracolsep{\fill}}rrrrr}
\hline\hline
& \multicolumn{1}{c}{$M_{bW,min}$} & \multicolumn{1}{c}{$M_{bW,max}$} 
& \multicolumn{1}{c}{$p_{Tb,min}$} & \multicolumn{1}{c}{$\Delta
R_{jj,min}$} & \multicolumn{1}{c}{$\Delta R_{jl,min}$} \\
\hline
Tevatron Run 1 & 150 GeV & 200 GeV & 35 GeV & 1.0 & 0.4 \\ 
Tevatron Run 2 & 150 GeV & 200 GeV & 35 GeV & 1.5 & 1.5 \\ 
Tevatron Run 3 & 150 GeV & 200 GeV & 35 GeV & 1.5 & 1.5 \\ 
LHC            & 145 GeV & 205 GeV & 35 GeV & 1.5 & 1.0 \\
\hline\hline
\end{tabular*}
\end{center}
\end{table}

\begin{table}
\caption{\label{ratesx} Cross sections in fb for the individual signal
(with $\kappa_f/\Lambda=0.2$~TeV$^{-1}$) and background channels after
the cuts in Eqs.~(\ref{basic1})--(\ref{basic3}) and in Table~\ref{cutsx}
are employed. Cross sections after $b$--tagging are shown in
parentheses. The signal values scale quadratically for $\kappa_f/\Lambda
\approxle 0.2$~TeV$^{-1}$.}
\vspace{0.2cm}
\begin{center}
\begin{tabular*}{\textwidth}{l@{\extracolsep{\fill}}rrrrrr}
\hline\hline
$\bar tcg$ & \multicolumn{2}{c}{Run 1} & \multicolumn{2}{c}{Runs 2 \& 3}
& \multicolumn{2}{c}{LHC} \\
\hline
$q\bar q \rightarrow t\bar c$ & 23 & (8) & 15 & (9) & 200 & (120)\\
$gg \rightarrow t\bar c$ & 87 & (32) & 76 & (45) & 15200 & (9090)\\
$cq \rightarrow tq$ & 56 & (20) & 52 & (31) & 4087 & (2450)\\
$cg \rightarrow tg$ & 130 & (47) & 115 & (69) & 21150 & (12650)\\
\hline
Total  & 296 & (107) & 258 & (154) & 40460 & (24310)\\
\hline\hline
$\bar tug$ & \multicolumn{2}{c}{Run 1} & \multicolumn{2}{c}{Runs 2 \& 3}
& \multicolumn{2}{c}{LHC} \\
\hline
$q\bar q \rightarrow t\bar u$ & 23 & (8) & 15 & (9) & 200 & (120)\\
$gg \rightarrow t\bar u$ & 87 & (32) & 76 & (45) & 15200 & (9090)\\
$uq \rightarrow tq$ & 1005 & (365) & 832 & (498) & 24760 & (14810)\\
$ug \rightarrow tg$ & 3025 & (1097) & 2381 & (1424) & 146200 & (87430) \\
\hline
Total  & 4140 & (1502) & 3304 & (1976) & 186360 & (111450)\\
\hline\hline
SM background & \multicolumn{2}{c}{Run 1} & \multicolumn{2}{c}{Runs 2 \& 3}
& \multicolumn{2}{c}{LHC} \\
\hline
$bq \rightarrow tq$  & 156 & (57) & 149 & (89) & 14570 & (8710)\\
$q\bar q \rightarrow t\bar b$ & 42 & (19) & 29 & (14) & 390 & (190)\\
$Wb\bar b$ & 62 & (29) & 13 & (6) & 160 & (80)\\
$Wjj$ & 7616 & (151) & 2270 & (45) & 97370 & (1930)\\
\hline
Total  & 7876 & (256) & 2461 & (154) & 112490 & (10910)\\
\hline \hline 
\end{tabular*}
\end{center}
\end{table}

\newpage

\begin{table}
\caption{\label{limitsx} The discovery limits on $\kappa_c / \Lambda$
and $\kappa_u / \Lambda$ for each of the collider options discussed in
the text. In both the charm and up quark cases, we assumed that the
coupling of the other type did not exist.}
\vspace{0.2 cm}
\begin{center}
\begin{tabular*}{\textwidth}{l@{\extracolsep{\fill}}rrrr}
\hline\hline
& \multicolumn{3}{c}{Tevatron} & \\ 
& \multicolumn{1}{c}{Run 1} & \multicolumn{1}{c}{Run 2} &
\multicolumn{1}{c}{Run 3} & \multicolumn{1}{c}{LHC} \\ \hline
$E_{cm}$ (TeV) & 1.8 & 2.0 & 2.0 & 14.0 \\ 
$\mathcal{L} \rm{~(fb}^{-1}\rm{)}$ & .1 & 2 & 30 & 10 \\ 
$\kappa_c / \Lambda \rm{~(TeV}^{-1}\rm{)}$ & .31 & .092 & .046 & .013 \\
$\kappa_u / \Lambda \rm{~(TeV}^{-1}\rm{)}$ & .082 & .026 & .013 & .0061
\\ \hline \hline 
\end{tabular*}
\end{center}
\end{table}

\newpage

\end{document}